# Design of an ultra-compact, energy-efficient non-volatile photonic switch based on phase change materials

Khoi Phuong Dao[1], Juejun Hu[1], and Richard Soref[2]

[1] Department of Materials Science and Engineering, Massachusetts Institute of Technology, Cambridge, MA 02139, USA

[2] Department of Engineering, University of Massachusetts Boston, Boston, Massachusetts 02125, USA

Corresponding author: Khoi Phuong Dao, email: khoidao@mit.edu.

**Abstract** The on-chip photonic switch is a critical building block for photonic integrated circuits (PICs) and the integration of phase change materials (PCMs) enables non-volatile switch designs that are compact, low-loss, and energy-efficient. Existing switch designs based on these materials typically rely on weak evanescent field interactions, resulting in devices with a large footprint and high energy consumption. Here we present a compact non-volatile 2 × 2 switch design leveraging optical concentration in slot waveguide modes to significantly enhance interactions of light with PCM, thereby realizing a compact, efficient photonic switch. To further improve the device's energy efficiency, we introduce an integrated single-layer graphene heater for ultrafast electrothermal switching of the PCM. Computational simulations demonstrate a 2 × 2 switch a crosstalk (CT) down to -24 dB at 1550 nm wavelength and more than 55 nm 0.3 dB insertion loss (IL) bandwidth. The proposed photonic switch architecture can constitute the cornerstone for next-generation high-performance reconfigurable photonic circuits.

## 1      Introduction

In the past few decades, the rapid development of PICs has demonstrated their potential in telecommunication and data communications. Moreover, the von Neumann bottleneck in electronics [1] suggests that scalable programmable PICs could be an alternative solution for energy-efficient classical and/or quantum information storage and processing[2,3]. Programmable PIC reported to date predominantly rely on thermo-optic effects [4], free-carrier effects [5], or electro-optic effects [6] of materials. The small change in refractive index afforded by these effects however limits the tunability and scalability of these methods, leading to a large device footprint and excessive energy consumption. Although plasmonic light confinement can mitigate the issue, that results in lossy devices [7], making the approach unsuitable for large-scale PICs. Moreover, these effects are volatile and demand a constant power supply (~10 mW). This disqualifies them for applications where only sporadic re-programming or reconfiguration is needed, such as optical switching and routing in data centers [8], optical neural networks [9], and photonic memories [10,11].

Chalcogenide-based phase change materials (PCMs) emerge as promising candidates to enable ultra-compact and energy-efficient reconfigurable photonics. They can reversibly switch between two stable states (amorphous and crystalline) in a non-volatile fashion and with exceptionally high refractive index contrast ($\Delta n \sim 1$) [11–14]. Phase transition in PCMs can be triggered by ultrashort optical or electrical pulses [15], and a multitude of intermediate states (between fully amorphous and crystalline) can be accessed by changing the pulse parameters [16]. In addition, PCMs offer compatibility with large scale integration, as they can be conveniently prepared using various large-area deposition methods [11,12,17,18] onto different photonic integrated circuit (PIC) platforms in a CMOS backend-compatible manner [12,19]. In spite of these advantages, conventional PCMs such as $Ge_2Se_2Te_5$ (GST) and GeTe display significant absorption in both phases at optical communication wavelengths, limiting their effectiveness in photonic phase shifters – a crucial component of programmable PICs. Recently, there have been growing interests in wide-bandgap PCMs such as GeSbSeTe (GSST) [20,21], antimony selenide ($Sb_2Se_3$) [22], and antimony sulfide ($Sb_2S_3$) [22,23]. For example, $Sb_2Se_3$ exhibits minimal losses at 1550 nm and a substantial index contrast ($\Delta n \approx 0.77$) [22]. These characteristics position $Sb_2Se_3$ as a promising phase-change material for applications in programmable photonics within the telecommunication bands.

One essential design element in PCM-based configurable devices is the heating mechanism. Electro-thermal heating using resistive micro-heaters facilitating scalable on-chip integration have been investigated in numerous recent studies. Various heater materials have been employed, including metals [24,25], transparent conducting oxides (TCOs) [26], and doped Si [27–29]. While metals prove effective for free-space reflective devices, they introduce notable optical losses in transmissive or waveguide components. Doped silicon stands out as an excellent choice for integrating PCMs into the silicon-on-insulator (SOI) platform. However, applying it to $Si_3N_4$-based devices, another widely used photonic platform, or to other non-silicon waveguide platforms poses challenges. TCO heaters, while suitable for devices operating in the visible spectrum, encounter exacerbated optical losses in the infrared due to free carrier absorption. To address these challenges, graphene has emerged as a promising heating material thanks to its exceptional thermal and electrical conductivity, versatile integration compatibilities, and remarkable stability [30,31]. Additionally, the infrared optical losses associated with graphene can be minimized by leveraging the doping-induced Pauli blocking effect. Recent theoretical analysis and experimental reports [32–34] indicates that graphene heaters exhibit two orders-of-magnitude higher figures of merit for overall performance (heating efficiency and induced loss) than that of doped Si or TCO heaters when applied to PCM switching.

Here, we present the design of a compact non-volatile photonic 2 × 2 switch on the SOI platform utilizing $Sb_2Se_3$ and a single-layer graphene heater. The design exploits a configuration involving a slotted waveguide filled with PCM. Compared to traditional layouts where the PCM typically interacts only with evanescent fields, the design leverages strong field concentration in the slot region to boost light-PCM interactions [35], thus, simultaneously achieves low insertion loss, a compact form factor, high extinction ratio, and zero-static power consumption.

## 2      Structure and design

Figure 1 shows the proposed 2 × 2 photonics directional-coupler switch design in a semi-standard SOI platform. The switch consists of a multimode slotted waveguide (the two-waveguide coupling zone) attached to four single-mode waveguides serving as input and output ports, on either side of the multimode section. The height and width of the single-mode waveguides are h = 240 nm and $W_{wg}$ = 450 nm, respectively. The slot waveguide has a length $L_{slot}$ = 10 µm and a centrally located slot with a width $W_{slot}$ = 100 nm, which is completely filled with $Sb_2Se_3$. In the telecommunication C-band, the refractive indices of $Sb_2Se_3$ are taken from [22] as 3.825 and 4.050 at 1550 nm wavelength for the amorphous $Sb_2Se_3$ and crystalline $Sb_2Se_3$ respectively. The loss of crystalline $Sb_2Se_3$ was also reported in the same paper to be as low as 0.01 dB $µm^{-1}$. The whole device is cladded all-around by $SiO_2$ with a thickness of 2 µm. Directly on top of the multimode slot waveguide, there is a single-layer graphene heater. The dimensions of the graphene layer are $L_{gr}$ = 9 µm and $W_{gr}$ = 3 µm and it is symmetrically positioned on the slot waveguide. Ti/Au contact pads are placed on both sides of the graphene heating layer to minimize the contact resistance. Figure 1b demonstrates the working principle of the switch as the PCM is switched reversibly by a sequence of voltage pulses applied to the graphene heater. Short pulses (several hundreds of nanoseconds) with high voltage will reset the PCM back to the amorphous state while the long pulses (several milliseconds) with moderate voltage are used to crystallize the material. $Sb_2Se_3$ was reported to be successfully amorphized at $T_a$ = 620 °C for 400 ns and crystallized at $T_c$ = 200 °C for minimally 0.1 ms [12]. In this study, the fundamental transverse electric (TE) mode at the telecommunication wavelength of 1550 nm was targeted. Yet, the design principle is not wavelength-sensitive and can be applied to a broadband device, as we will show later. The whole design could be realized by standard lithography and dry etching processes. $Sb_2Se_3$ be deposited in the slot by conformal coating methods such as atomic layer deposition [36] and solution processing [37]. Alternatively, thermal reflow has been demonstrated as an effective means for filling thin slots with chalcogenide materials [38]. The graphene heater can be fabricated by transferring chemical-vapor-deposition (CVD) grown single-layer graphene via the standard wet transfer technique [39], followed by lithographic patterning and metallization.

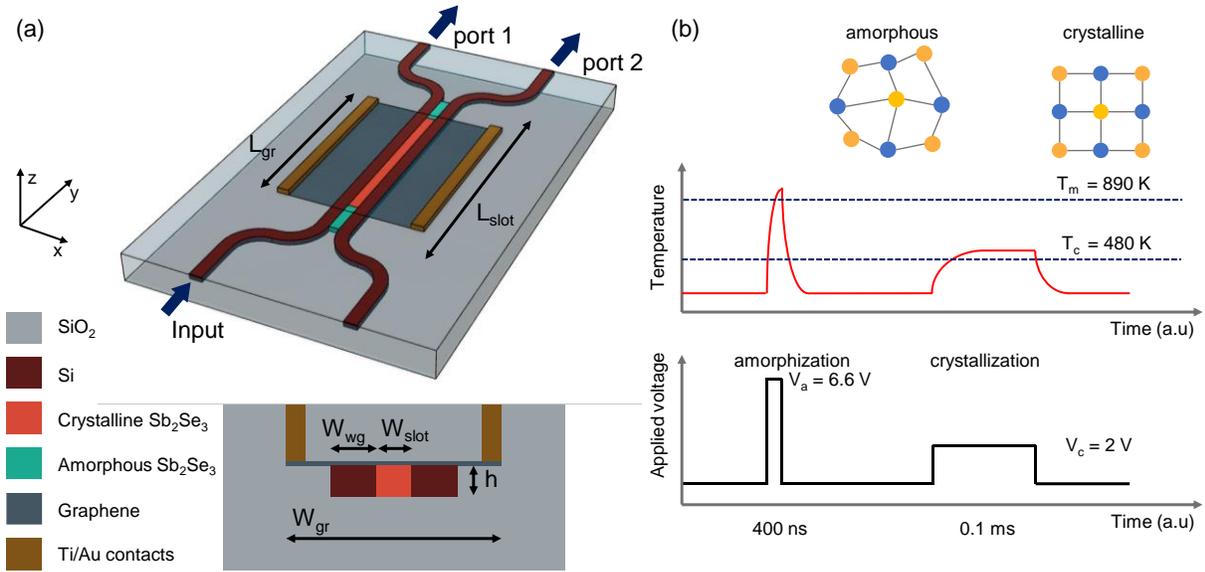

**Fig. 1.** (a) Schematic of the 2 × 2 photonic switch based on $Sb_2Se_3$. (b) Schematic of switching operation utilizing amorphization/crystallization voltage pulses to induce Joule heating.

The substantial refractive index contrast provided by $Sb_2Se_3$ facilitates the generation of even and odd TE modes within the multimode slot waveguide region, characterized by significant shifts in effective indices upon transitioning the phase of the slot material (as depicted in Figure 2(a), and 2(b)). Exploiting the disparities in the confinement and effective indices of these supermodes at amorphous and crystalline states, the photonic switch can be dynamically shifted from a bar state to a cross state. Due to the difference in propagating constants, the interference between odd mode and even mode results in the oscillation between bar state and cross state when we increase the slot length. The ultralow loss exhibited by $Sb_2Se_3$ allows the PCM to have strong overlap with the supermodes in the slot region without incurring excessive losses, thereby enhancing the phase modulation effect. The effective index differences between the even mode and the odd mode are 0.2951 and 0.3875 for amorphous and crystalline $Sb_2Se_3$ states, respectively, corresponding to different beating lengths in the two states.

Figure 2(c) plots transmission through the ports 1 and 2 in the amorphous $Sb_2Se_3$ state with different PCM-filled slot length $L_{slot}$, simulated using 3D finite-difference time-domain (FDTD) and fitted to sine curves. Following the result, we take the slot length to be 10 µm, which yields the maximum transmission in the amorphous (bar) state. To realize switching with maximum contrast, we selectively crystallize the center section of the PCM as illustrated in Figure. 2(d). In practice, this can be implemented by controlling the temperature profile along the y-axis, as the center portion of the PCM slot experiences the highest temperature and hence preferentially crystallizes first (refer to Fig. 4 and discussions for more details). Figure 2(d) plots the simulated transmission through the output ports as a function of crystallized $Sb_2Se_3$ slot length. The plot implies that an optical crystallization section length of 8.5 µm (i.e., leaving 0.75 µm of amorphous region on either side) would result in maximal switching contrast. This leads to an overall device (including the four single-mode waveguide ports) footprint of 5.5 µm x 24 µm. We note that partial crystallization of PCMs has been discussed in a number of literature reports showing good reproducibility [40–42]. The main difference between our assumption here and actual experiment implementation is that there is no abrupt interface between the amorphous and crystalline $Sb_2Se_3$ sections. Instead, a gradual transition region with varying crystalline fraction is likely present. This contributes to lowering reflection and scattering from the abrupt interface, and can lead to even lower insertion losses than simulations presented here.

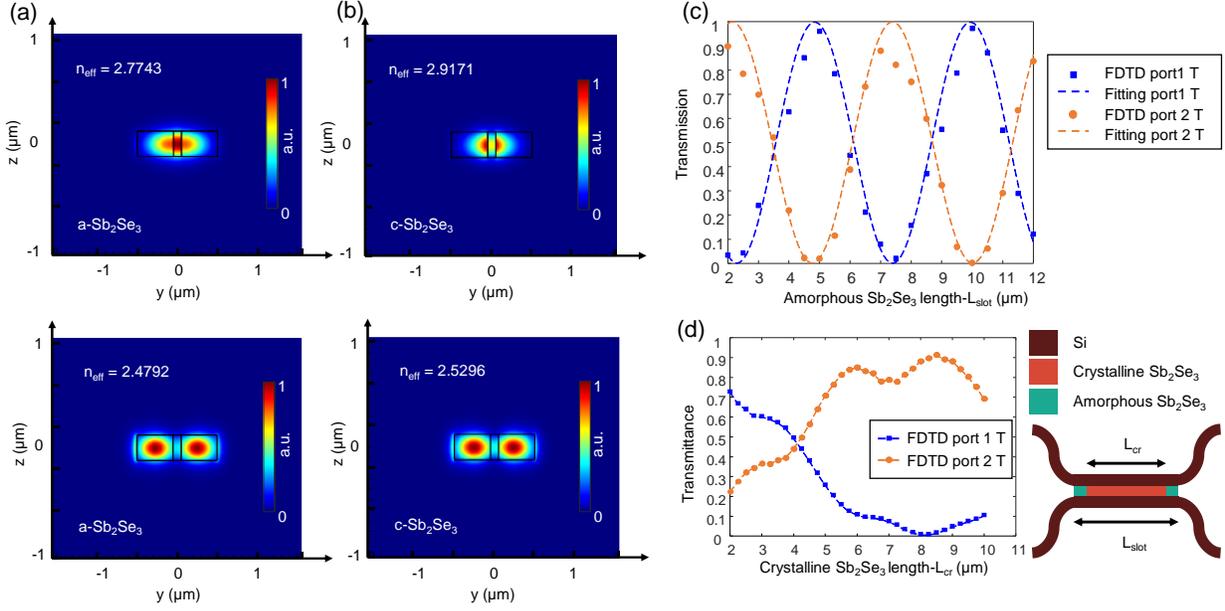

**Fig. 2.** (a), (b) The cross-sectional electric field distribution for even and odd TE modes of the multimode slot waveguide, (c) Transmission through the two ports as functions of PCM length, (d) Transmission through the two ports as function of crystallized PCM section's length.

## 3     Results

We conducted 3D FDTD simulations to validate the switching efficiency of our proposed design. Figures 3(a) and 3(b) depict the transmission spectra of our $2 \times 2$ photonic switch and the corresponding in-plane electric field distributions corresponding to amorphous and crystalline $Sb_2Se_3$ states. The overall insertion loss (IL) is -0.27 dB (cross state) and -0.11 dB (bar state) at 1550 nm, and is consistently less than 0.5 dB across 1525 nm to 1575 nm wavelengths. The 0.3 dB IL bandwidth is no less than 55 nm. The crosstalk (CT), defined as contrast ratio between the on/off states at the output ports, reaches -23.9 dB (cross state) and -27.4 dB (bar state) at 1550 nm, and stays better than -15 dB throughout the 1525 nm to 1575 nm band. These performance metrics compare favorably to state-of-the-art PCM switches as summarized in Table 1.

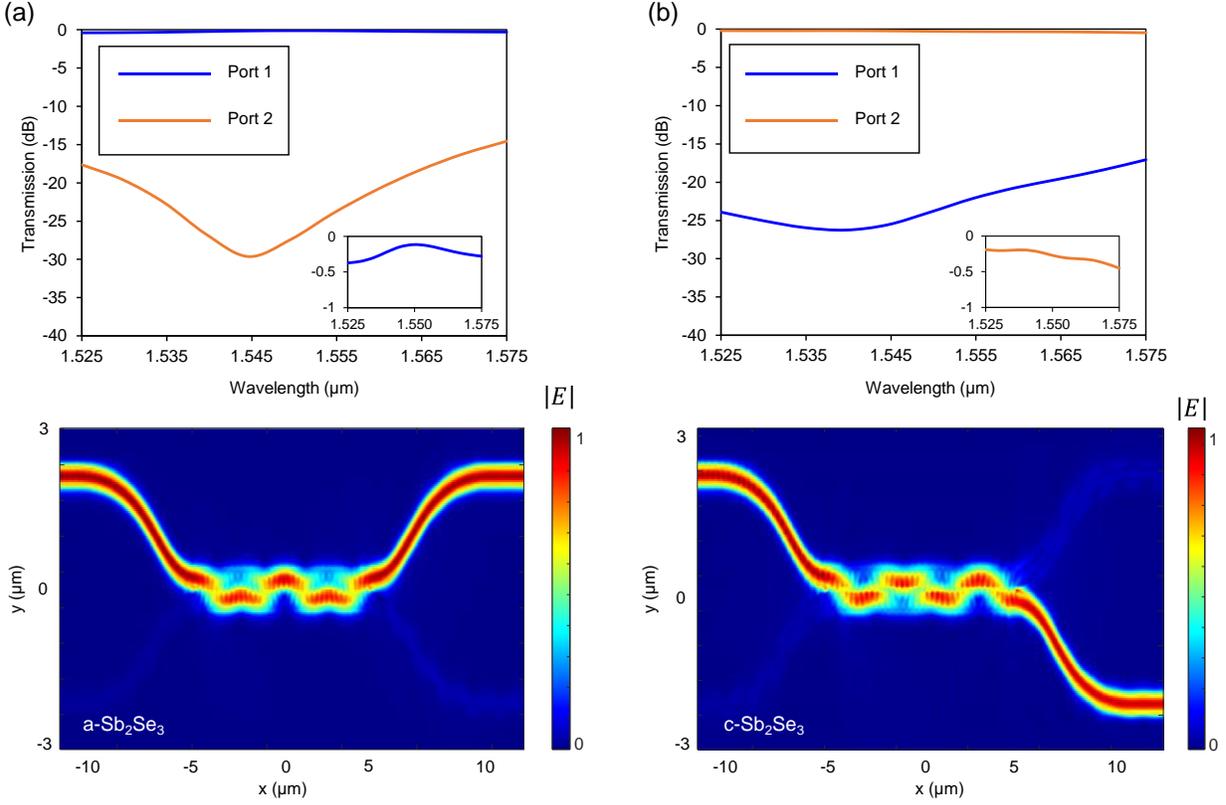

**Fig. 3.** Transmission spectra at two output port and the normalized electric field intensity distribution of the 2 × 2 switch at 1550 nm with (a) amorphous $Sb_2Se_3$ (bar state) and (b) crystalline $Sb_2Se_3$ (cross state)

Next, we discuss the thermal performance of the single-layer graphene heater. The graphene heater has been reported to have outstanding performance as a heating element for integrated photonic devices incorporating PCMs, offering exceptional energy efficiency and high operational speed [31,32], owing to its ultralow heat capacity and high in-plane thermal and electrical conductivity. Compared to doped Si, which is another popular choice of heater material, graphene heaters claim significantly lower induced loss and higher heating efficiency [32,33]. Joule heating employing the graphene heater follows similar phase change dynamics demonstrated in [12,29]. The pulse width and voltage are contingent on the microheater's properties. With our specific proposed graphene heater, pulses of 2V, which induces a current of 5.67 mA, with duration of 100 µs are applied to partially crystallize the $Sb_2Se_3$ slot, heating its 8.5 µm long center section to above the crystallization temperature $T_c$ (here set as 200 °C). It is reported that pulses (as short as 5 µs) can crystallize $Sb_2Se_3$ but result in spatially non-uniform crystallization. Longer pulses lasting 100 µs or more are necessary to crystallize the PCM uniformly [12], which justifies our pulse parameter choice here. To induce amorphization, we investigated two types of pulses, a single 8.6 V (22.4 mA), 100 ns pulse [12,22] or a 6.6 V (18.7 mA), 400 ns pulse. Both can elevate the temperature of the entire PCM-filled slot above the melting point, $T_m$ = 620 °C (893 K). The total energy consumption for crystallization is 1.13 µJ and 21/49.4 nJ (100/400 ns pulse) for amorphization. Clearly, a trade-off between pulse voltage and switch energy exists for the amorphization process. Figure 4(a) demonstrates the temperature evolution over amorphization and crystallization cycles from finite-element method (FEM) simulations using COMSOL Multiphysics. For crystallization, the temperature remains stably higher than $T_c$ across the target section. Figure 4(b) plots the temperature across the $Sb_2Se_3$ slot at the end of the crystallization and amorphization (100 ns) pulses. The coordinate z = 0 nm refer to the boundary between the silicon waveguide layer and the buried oxide (BOX). The yellow line shows that the entire $Sb_2Se_3$ middle section (from y = -4.25 µm to y = 4.25 µm) was elevated to be above its melting point, which guarantees complete amorphization. The 3D temperature profile predicted by thermal FEM simulation at the end of the 100 ns amorphization pulse was shown in Figure 4(c), suggesting that the heat is effectively confined within the target section of $Sb_2Se_3$.

The out-of-plane temperature variation is particularly relevant for graphene heaters, given that graphene exhibits varying out-of-plane thermal resistance due to the surface polar phonons (SPoPh) scattering effect [43]. Consequently, a temperature gradient is established along the out-of-plane direction, as heat transfer occurs more efficiently towards the substrate than towards the top cladding. In order to ensure complete amorphization of $Sb_2Se_3$ following the melt-quenching pulse, a series of dynamic simulations was conducted with varying amorphization pulse power. Thermal FEM simulations (Figure 4(b)) suggest that the crystallization length within the slot barely changes at z = 0 and z = 240 nm, implying that the crystallization is uniform along the z-direction. The kinks near the two ends of the orange curve in Figure 4(b) are attributed to the ends of the graphene sheet.

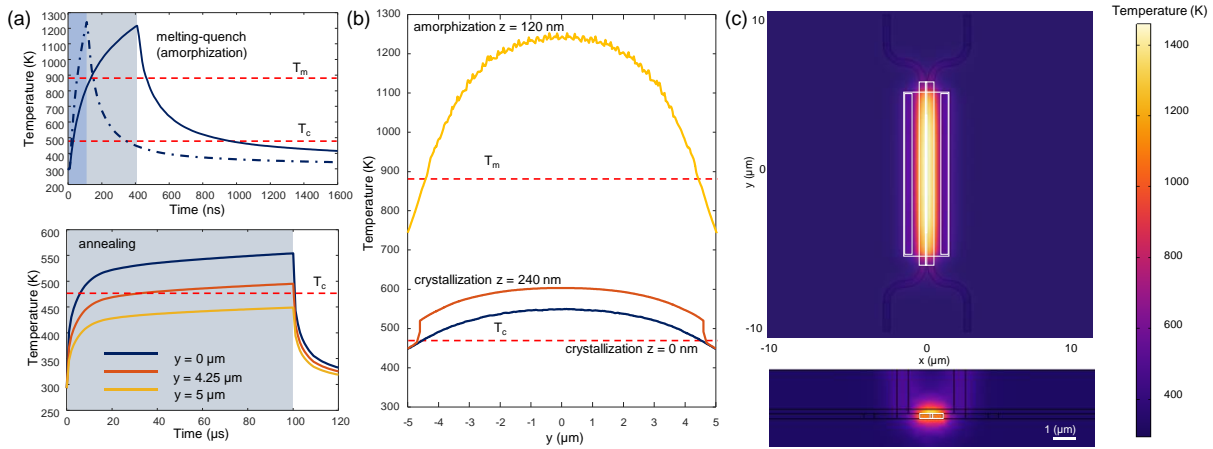

**Fig 4**. (a) FEM simulated transient temperature during and after the amorphization and crystallization pulses. The shaded areas represent pulse-on times. (b) Simulated temperature across the slot at the end of the amorphization and crystallization pulses. (c) Simulated three-dimension temperature profiles at the end of the amorphization pulse.

The quenching rate after the amorphization pulse, which is critical in gauging whether crystallization can be bypassed, is predominantly governed by thermal conductance through cladding material and BOX. As shown in Figure 4(a), the quenching rate is approximately 1 K/ns, which is sufficient to prohibit re-crystallization of $Sb_2Se_3$.

Table 1. Comparison of PCM-based optical switches

| Design | Ref | PCM | IL (dB) | CT (dB) | Footprint (μm²) | Optical BW (nm) | Switching energy |
|---|---|---|---|---|---|---|---|
| DC* | [14] | GST | 2 | -10 | 5 × 45 | 30 | - |
| DC* | [18] | GST | 2 | -10 | 5 × 50 | 30 | 380 nJ (6.8 μJ) |
| MZI* | [12] | $Sb_2Se_3$ | 0.3 [a] | 6.5/15 | 100 × 100 | 15 | 176 nJ (3.8 μJ) |
| MZI* | [44] | $Sb_2S_3$ | 3 | -12 | 100 × 100 | 20 | - |
| MRR | [45] | GST | 2 | -20 | 15 × 15 | <1 | - |
| MRR* | [46] | GST | 3 | 14 | 25 × 25 | <1 | 0.25 nJ (11 nJ) [c] |
| MRR* | [47] | GST | 5.1/4.3 [b] | 5 | 60 × 60 | <1 | 0.19 nJ (17.1 nJ) [c] |
| MMI* | [48] | $Sb_2Se_3$ | 0.5 [a] | 8 | 6 × 33 | - | 14 nJ (0.95 mJ) [c] |
| DC | [49] | $Sb_2S_3$ | 0.26 | -31.3 | 4.9 × 25.4 | 35 | 9.59nJ (-) |
| DC | This work | $Sb_2Se_3$ | 0.3 | -23.9 | 5.5 × 24 | 55 | 21 nJ (1.13μJ) |

DC: directional coupler, MZI: Mach-Zehnder Interferometer, MRR: micro-ring resonator, MMI: Multimode Interferometer, IL: insertion loss, CT: crosstalk, Optical BW: optical bandwidth for the corresponding IL, Switching energy: Energy per switching event for amorphization/(crystallization)

*Experimental results

[a] additional loss due to PCMs to the total device IL, [b] through/drop port IL, [c] optical pulse energy

Finally, we assess the scalability of our design to large switch matrices. $2^m \times 2^m$ switches built from $2 \times 2$ building blocks using the Benes network can be used to estimate the performance of our proposed design [20]. Using the values presented in Table 1, we can estimate the total insertion loss and the lower and maximum crosstalk of an m-order switch matrix (assuming that the IL of a waveguide crossing in the C-band as 0.1 dB [50]):

$$IL_m = (2^m - 2) \times 0.1 \, dB + (2m - 1) \times 0.45 \, dB \tag{1}$$

$$CT_m = -(15 \, dB - 10 \log_{10} m \, dB) \tag{2}$$

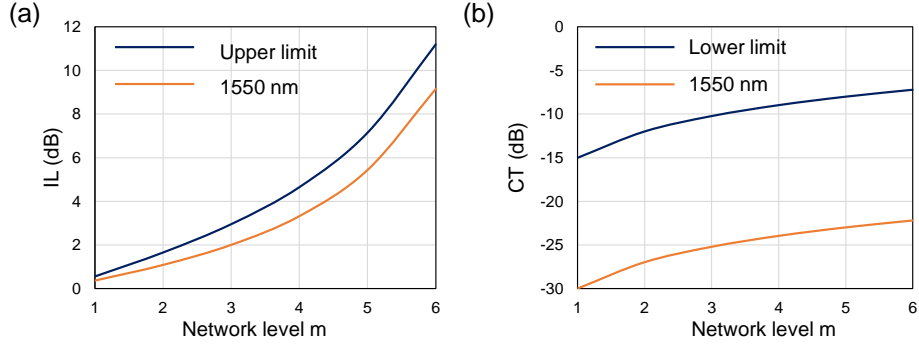

**Fig. 5.** (a) Insertion loss and (b) maximum crosstalk as functions of the switch matrix of order m

Scaling from our 2 × 2 switch's performance, a 16 × 16 switch is anticipated to exhibit maximal 3.2 dB insertion loss and -24 dB crosstalk at 1550 nm, representing highly promising performance metrics, compared to the state-of-the-art (volatile) on-chip 16 × 16 switch reported by Lu et al [51]. The devices can also be put to good use in creating large-scale, programmable, rectangular and triangular and hexagonal meshes [52].

## 4 Conclusion

In conclusion, the non-volatile 2 × 2 photonic switch design takes advantage of a PCM-slot configuration to achieve an ultra-compact footprint (5.5 × 24 µm$^2$) with minimal crosstalk (-23.9 dB), and the low-loss PCM $Sb_2Se_3$ enables a low insertion loss of 0.27 dB, and a single-layer graphene heater achieves low switching energies of 1.13 µJ for crystallization and 21 nJ for amorphization. The design further demonstrates its scalability toward large-scale non-blocking switch matrices. Our proposed design, therefore, holds the potential for the development of future large-scale PCM-based programmable PICs.

*Disclosure*

The authors declare no conflict of interests

*Data availability statement*

The data that support the findings of this study are available from the corresponding author, KPD, upon reasonable request.

*Acknowledgements*

Funding support is provided by NSF under award number 2132929 and the Air Force Office of Scientific Research AFOSR under grant numbers FA9550-21-1-0347 and FA9550-22-1-0532. We acknowledge the insightful discussions and technical assistance from Professor Francesco De Leonardis.